\documentclass[%
reprint,
superscriptaddress,
showpacs,
 amsmath,amssymb,
 aps,prl,
floatfix,
]{revtex4-2}

\usepackage[english]{babel}
\usepackage[utf8]{inputenc}
\usepackage{xcolor}
\usepackage{amsthm}
\usepackage{textcomp}
\usepackage{mathtools}
\usepackage{multirow}
\usepackage{graphicx}
\usepackage{adjustbox}
\usepackage{placeins}
\usepackage[T1]{fontenc}
\usepackage{lipsum}
\usepackage{csquotes}
\usepackage[pdftex, pdftitle={Article}, pdfauthor={Author},colorlinks = true,allcolors = blue]{hyperref} 
\usepackage{dcolumn}
\usepackage{bm}
\usepackage{comment}
\usepackage{blindtext}
\usepackage{hyperref}
\hypersetup{
     colorlinks   = true,
     citecolor    = blue,
     linkcolor    = blue,
     urlcolor     = blue
}

\usepackage{etoolbox}

\usepackage[mathlines]{lineno}

\begin{document}

\title{First Measurement of $\Lambda$ Electroproduction off Nuclei in the Current and Target Fragmentation Regions}

\newcommand*{\ANL}{Argonne National Laboratory, Argonne, IL 60439}
\newcommand*{\ANLindex}{1}
\affiliation{\ANL}
\newcommand*{\CSUDH}{California State University, Dominguez Hills, Carson, CA 90747}
\newcommand*{\CSUDHindex}{2}
\affiliation{\CSUDH}
\newcommand*{\CANISIUS}{Canisius College, Buffalo, NY 14208}
\newcommand*{\CANISIUSindex}{3}
\affiliation{\CANISIUS}
\newcommand*{\CMU}{Carnegie Mellon University, Pittsburgh, PA 15213}
\newcommand*{\CMUindex}{4}
\affiliation{\CMU}
\newcommand*{\CUA}{Catholic University of America, Washington, D.C. 20064}
\newcommand*{\CUAindex}{5}
\affiliation{\CUA}
\newcommand*{\SACLAY}{IRFU, CEA, Universit\'{e} Paris-Saclay, F-91191 Gif-sur-Yvette, France}
\newcommand*{\SACLAYindex}{6}
\affiliation{\SACLAY}
\newcommand*{\CNU}{Christopher Newport University, Newport News, VA 23606}
\newcommand*{\CNUindex}{7}
\affiliation{\CNU}
\newcommand*{\UCONN}{University of Connecticut, Storrs, CT 06269}
\newcommand*{\UCONNindex}{8}
\affiliation{\UCONN}
\newcommand*{\DUKE}{Duke University, Durham, NC 27708-0305}
\newcommand*{\DUKEindex}{9}
\affiliation{\DUKE}
\newcommand*{\DUQUESNE}{Duquesne University, 600 Forbes Avenue, Pittsburgh, PA 15282 }
\newcommand*{\DUQUESNEindex}{10}
\affiliation{\DUQUESNE}
\newcommand*{\FU}{Fairfield University, Fairfield CT 06824}
\newcommand*{\FUindex}{11}
\affiliation{\FU}
\newcommand*{\FERRARAU}{Universit\`{a} di Ferrara, 44121 Ferrara, Italy}
\newcommand*{\FERRARAUindex}{12}
\affiliation{\FERRARAU}
\newcommand*{\FIU}{Florida International University, Miami, FL 33199}
\newcommand*{\FIUindex}{13}
\affiliation{\FIU}
\newcommand*{\FSU}{Florida State University, Tallahassee, FL 32306}
\newcommand*{\FSUindex}{14}
\affiliation{\FSU}
\newcommand*{\GWUI}{The George Washington University, Washington, D.C. 20052}
\newcommand*{\GWUIindex}{15}
\affiliation{\GWUI}
\newcommand*{\GSIFFN}{GSI Helmholtzzentrum fur Schwerionenforschung GmbH, D-64291 Darmstadt, Germany}
\newcommand*{\GSIFFNindex}{16}
\affiliation{\GSIFFN}
\newcommand*{\INFNFE}{INFN, Sezione di Ferrara, 44100 Ferrara, Italy}
\newcommand*{\INFNFEindex}{17}
\affiliation{\INFNFE}
\newcommand*{\INFNFR}{INFN, Laboratori Nazionali di Frascati, 00044 Frascati, Italy}
\newcommand*{\INFNFRindex}{18}
\affiliation{\INFNFR}
\newcommand*{\INFNGE}{INFN, Sezione di Genova, 16146 Genova, Italy}
\newcommand*{\INFNGEindex}{19}
\affiliation{\INFNGE}
\newcommand*{\INFNRO}{INFN, Sezione di Roma Tor Vergata, 00133 Rome, Italy}
\newcommand*{\INFNROindex}{20}
\affiliation{\INFNRO}
\newcommand*{\INFNTUR}{INFN, Sezione di Torino, 10125 Torino, Italy}
\newcommand*{\INFNTURindex}{21}
\affiliation{\INFNTUR}
\newcommand*{\INFNPAV}{INFN, Sezione di Pavia, 27100 Pavia, Italy}
\newcommand*{\INFNPAVindex}{22}
\affiliation{\INFNPAV}
\newcommand*{\ORSAY}{Universit\'{e} Paris-Saclay, CNRS/IN2P3, IJCLab, 91405 Orsay, France}
\newcommand*{\ORSAYindex}{23}
\affiliation{\ORSAY}
\newcommand*{\Juelich}{Institut f\"ur Kernphysik (Juelich), Juelich, 52428, Germany}
\newcommand*{\Juelichindex}{24}
\affiliation{\Juelich}
\newcommand*{\JMU}{James Madison University, Harrisonburg, VA 22807}
\newcommand*{\JMUindex}{25}
\affiliation{\JMU}
\newcommand*{\KNU}{Kyungpook National University, Daegu 41566, Republic of Korea}
\newcommand*{\KNUindex}{26}
\affiliation{\KNU}
\newcommand*{\LAMAR}{Lamar University, 4400 MLK Blvd, PO Box 10046, Beaumont, TX 77710}
\newcommand*{\LAMARindex}{27}
\affiliation{\LAMAR}
\newcommand*{\MIT}{Massachusetts Institute of Technology, Cambridge, MA  02139-4307}
\newcommand*{\MITindex}{28}
\affiliation{\MIT}
\newcommand*{\MISS}{Mississippi State University, Mississippi State, MS 39762-5167}
\newcommand*{\MISSindex}{29}
\affiliation{\MISS}
\newcommand*{\ITEP}{National Research Centre Kurchatov Institute - ITEP, Moscow, 117259, Russia}
\newcommand*{\ITEPindex}{30}
\affiliation{\ITEP}
\newcommand*{\UNH}{University of New Hampshire, Durham, NH 03824-3568}
\newcommand*{\UNHindex}{31}
\affiliation{\UNH}
\newcommand*{\NMSU}{New Mexico State University, PO Box 30001, Las Cruces, NM 88003, USA}
\newcommand*{\NMSUindex}{32}
\affiliation{\NMSU}
\newcommand*{\NSU}{Norfolk State University, Norfolk, VA 23504}
\newcommand*{\NSUindex}{33}
\affiliation{\NSU}
\newcommand*{\OHIOU}{Ohio University, Athens, OH  45701}
\newcommand*{\OHIOUindex}{34}
\affiliation{\OHIOU}
\newcommand*{\ODU}{Old Dominion University, Norfolk, VA 23529}
\newcommand*{\ODUindex}{35}
\affiliation{\ODU}
\newcommand*{\JLUGiessen}{II Physikalisches Institut der Universitaet Giessen, 35392 Giessen, Germany}
\newcommand*{\JLUGiessenindex}{36}
\affiliation{\JLUGiessen}
\newcommand*{\RPI}{Rensselaer Polytechnic Institute, Troy, NY 12180-3590}
\newcommand*{\RPIindex}{37}
\affiliation{\RPI}
\newcommand*{\URICH}{University of Richmond, Richmond, VA 23173}
\newcommand*{\URICHindex}{38}
\affiliation{\URICH}
\newcommand*{\ROMAII}{Universit\'{a} di Roma Tor Vergata, 00133 Rome Italy}
\newcommand*{\ROMAIIindex}{39}
\affiliation{\ROMAII}
\newcommand*{\MSU}{Skobeltsyn Institute of Nuclear Physics, Lomonosov Moscow State University, 119234 Moscow, Russia}
\newcommand*{\MSUindex}{40}
\affiliation{\MSU}
\newcommand*{\SCAROLINA}{University of South Carolina, Columbia, SC 29208}
\newcommand*{\SCAROLINAindex}{41}
\affiliation{\SCAROLINA}
\newcommand*{\TEMPLE}{Temple University,  Philadelphia, PA 19122 }
\newcommand*{\TEMPLEindex}{42}
\affiliation{\TEMPLE}
\newcommand*{\JLAB}{Thomas Jefferson National Accelerator Facility, Newport News, VA 23606}
\newcommand*{\JLABindex}{43}
\affiliation{\JLAB}
\newcommand*{\UTFSM}{Universidad T\'{e}cnica Federico Santa Mar\'{i}a, Casilla, 110-V Valpara\'{i}so, Chile}
\newcommand*{\UTFSMindex}{44}
\affiliation{\UTFSM}
\newcommand*{\CCTVal}{Center for Science and Technology of Valpara\'iso 699, Valpara\'iso, Chile}
\newcommand*{\CCTValindex}{34}
\affiliation{\CCTVal}
\newcommand*{\SAPHIR}{SAPHIR Millennium Science Institute, Santiago, Chile}
\newcommand*{\SAPHIRindex}{34}
\affiliation{\SAPHIR}
\newcommand*{\BRESCIA}{Universit\`{a} degli Studi di Brescia, 25123 Brescia, Italy}
\newcommand*{\BRESCIAindex}{45}
\affiliation{\BRESCIA}
\newcommand*{\UCR}{University of California Riverside, 900 University Avenue, Riverside, CA 92521, USA}
\newcommand*{\UCRindex}{46}
\affiliation{\UCR}
\newcommand*{\GLASGOW}{University of Glasgow, Glasgow G12 8QQ, United Kingdom}
\newcommand*{\GLASGOWindex}{47}
\affiliation{\GLASGOW}
\newcommand*{\YORK}{University of York, York YO10 5DD, United Kingdom}
\newcommand*{\YORKindex}{48}
\affiliation{\YORK}
\newcommand*{\VT}{Virginia Tech, Blacksburg, VA 24061-0435}
\newcommand*{\VTindex}{49}
\affiliation{\VT}
\newcommand*{\VIRGINIA}{University of Virginia, Charlottesville, VA 22901}
\newcommand*{\VIRGINIAindex}{50}
\affiliation{\VIRGINIA}
\newcommand*{\WM}{College of William and Mary, Williamsburg, VA 23187-8795}
\newcommand*{\WMindex}{51}
\affiliation{\WM}
\newcommand*{\YEREVAN}{Yerevan Physics Institute, 375036 Yerevan, Armenia}
\newcommand*{\YEREVANindex}{52}
\affiliation{\YEREVAN}

\newcommand*{\NOWISU}{Idaho State University, Pocatello, Idaho 83209}

\author {T.~Chetry} 
\affiliation{\MISS}
\affiliation{\FIU}
\author{L.~El~Fassi}\email{le334@msstate.edu}
\affiliation{\MISS}
\author {W.~K.~Brooks} 
\affiliation{\UTFSM}
\affiliation{\CCTVal}
\affiliation{\SAPHIR}
\affiliation{\JLAB}
\author {R.~Dupr\'e} 
\affiliation{\ORSAY}
\author {A.~El~Alaoui} 
\affiliation{\UTFSM}
\author {K.~Hafidi} 
\affiliation{\ANL}
\author {P.~Achenbach} 
\affiliation{\JLAB}
\author {K.P.~Adhikari} 
\affiliation{\MISS}
\author {Z.~Akbar} 
\affiliation{\VIRGINIA}
\author {W.R. Armstrong} 
\affiliation{\ANL}
% ...
\collaboration{The CLAS Collaboration}
\noaffiliation

\date{\today}

\begin{abstract}

We report results of $\Lambda$ hyperon production in semi-inclusive deep-inelastic scattering off deuterium, carbon, iron, and lead targets obtained with the CLAS detector and the Continuous Electron Beam Accelerator Facility 5.014~GeV electron beam. These results represent the first measurements of the $\Lambda$ multiplicity ratio and transverse momentum broadening as a function of the energy fraction~($z$) in the current and target fragmentation regions. The multiplicity ratio exhibits a strong suppression at high~$z$~and~an enhancement at~low~$z$. The measured transverse momentum broadening is an order of magnitude greater than that seen for light mesons. This indicates that the propagating entity interacts very strongly with the nuclear medium, which suggests that propagation of diquark configurations in the nuclear medium takes place at least part of the time, even at high~$z$. The trends of these results are qualitatively described by the Giessen Boltzmann-Uehling-Uhlenbeck transport model, particularly for the multiplicity ratios. These observations will potentially open a new era of studies of the structure of the nucleon as well as of strange baryons.

\end{abstract}

\maketitle

\begin{figure}[t]
\centering
\includegraphics[clip=true, trim= 0 0 0 2.75cm,width=0.42\textwidth, keepaspectratio=TRUE]{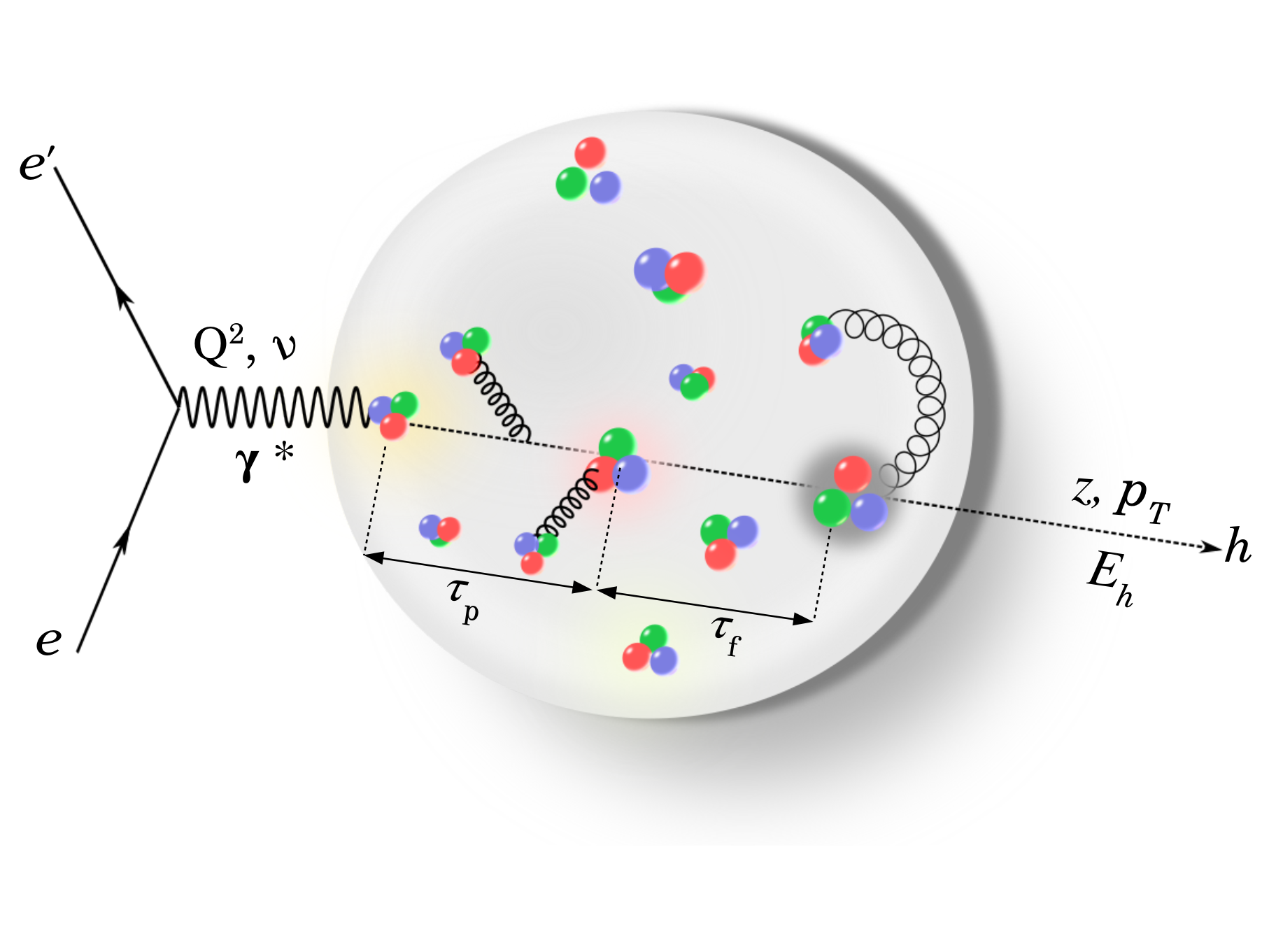}
\vglue -0.5in
\caption{An illustration of the hadronization process as well as its production, $\tau_p$, and formation, $\tau_f$, time-scales. $\nu= E_{e}-E_{e'}$ is the $\gamma^*$ energy transferred to the struck quark, $Q^{2}$ is the four-momentum transfer squared, $z= E_{h}/\nu$ is the fractional energy of the observed hadron, $h$, where $E_h$ is the hadron's energy in the lab frame, and $p_T$ is the hadron's transverse momentum with respect to the virtual-photon direction (see Fig.~\ref{fig:mass_spect} top right).}
\label{fig:hadproc}
\end{figure}

The study of the underlying structure of hadrons suggests a dynamical origin of the strong interactions between the confined color objects, quarks and gluons (partons), the building blocks of nuclei. Given that the description of the nonperturbative transition from partonic degrees of freedom to ordinary hadrons cannot be performed within the perturbative quantum chromodynamics (QCD) or lattice QCD frameworks, pure phenomenological methods are explored to study low-energy phenomena such as the hadronization process~\cite{GrossWilczek73, Dokshitzer:2003bt}.~To this end, deep-inelastic electron-nucleon scattering (DIS) has been utilized as a pioneering process on atomic nuclei to access the modified parton distributions, test the hadronization mechanisms, and study color confinement dynamics in the cold nuclear medium~\cite{Osborne:1978ai, EuropeanMuon:1991jmx, EuropeanMuon:1984inj}.~In this regime, when the electron emits an energetic virtual-photon ($\gamma^*$) that removes the struck quark from the rest of the residual system, it takes a finite time until the reaction products hadronize. These products would, in lepton-nucleus scattering, interact with the surrounding nuclear medium during the formation time, which is approximated at intermediate energies to be of a similar order as nuclear radii~\cite{Gallmeister:2007an}. The target nucleus acts then as a femtoscope with unique analyzing power that allows for the extraction of the hadronization time-distance scales. Therefore, the study of scattering off nuclei with different sizes and at various $\gamma^*$ kinematics probes the space-time evolution of the hadronization mechanism related to the quark propagation and the color field restoration to form regular hadrons~\cite{Accardi:2009qv,BROOKS2021136171}.\newline
\indent
As depicted in Fig.~\ref{fig:hadproc}, the hadronization process is characterized by two timescales describing its two phases. After the virtual photon hard scattering, during the production time~($\tau_{p}$), the struck quark propagates in the medium as a colored object and thus emits gluons (even in vacuum).~This quark then transforms into a colorless object, referred to as a prehadron, which eventually evolves into a fully dressed hadron within the formation time~($\tau_f$). The hadronization studies are thus performed to provide information on the dynamics scales of the process, and to constrain the existing models that provide different predictions of its time characteristics either in vacuum or in nuclei~\cite{Andersson:1979ue, Artru:1974hr, Shuryak:1978ij, Wang:2003aw, Kopeliovich:2003py}.~In principle, the production and formation mechanisms are the same for both cases with the exception that in the former, the~$q\bar{q}$ pairs or $qqq$ systems are considered emerging from the vacuum before expanding into color singlet hadrons, while in the latter, the struck quark is propagating and can pick up its partner(s) from the medium.~In this case, the presence of the medium will lead to several modifications and in-medium stimulated effects related either to the struck quark, formed prehadron, and/or hadron interactions with their surroundings.\\
\indent
The study of hadronization mechanisms is done in the framework of semi-inclusive DIS (SIDIS), and its characteristics are probed via the measurement of two experimental observables. The first is the hadron multiplicity ratio, $R^{A}_{h}$, which is defined as
\begin{equation}
\label{MReq}
 R^{A}_{h}(\nu,Q^2,z,p_{T}^2)= \frac{N^{A}_{h}(\nu,Q^2,z,p_T^2)/N^{A}_{e}(\nu,Q^2)}{N^{D}_{h}(\nu,Q^2,z,p_T^2)/N^{D}_{e}(\nu,Q^2)},
\end{equation}\noindent
where $N^{A}_{e}$ and $N^{A}_{h}$ are, respectively, the scattered electron and SIDIS hadron yields produced on a target~$A$ and corrected for detector acceptance and reconstruction efficiency.~The variables $\nu$, $Q^{2}$, $z$, and $p_{T}$ are defined in Fig.~\ref{fig:hadproc}.~The multiplicity ratio is normalized by DIS electrons originating from corresponding targets to cancel, to some extent, the initial-state nuclear effects and thus correct for the European Muon Collaboration (EMC) effect~\cite{Accardi:2009qv}. $R^A_h$ quantifies to which extent hadrons are attenuated at a given kinematics as was reported in earlier studies by SLAC~\cite{Osborne:1978ai}, HERMES~\cite{Airapetian:2000ks,Airapetian:2003mi,Airapetian:2007vu,Airapetian:2009jy,Airapetian:2011jp}, and EMC~\cite{EuropeanMuon:1991jmx} due to the (pre)hadron elastic or inelastic scattering and/or the energy loss of the hadron-fragmented struck quark during the color-neutralization stage preceding hadron formation.

\begin{figure}[b]
\includegraphics[clip=true, trim= 0.25cm 0.02cm 0 0.275cm,width=0.5\textwidth, keepaspectratio=TRUE]{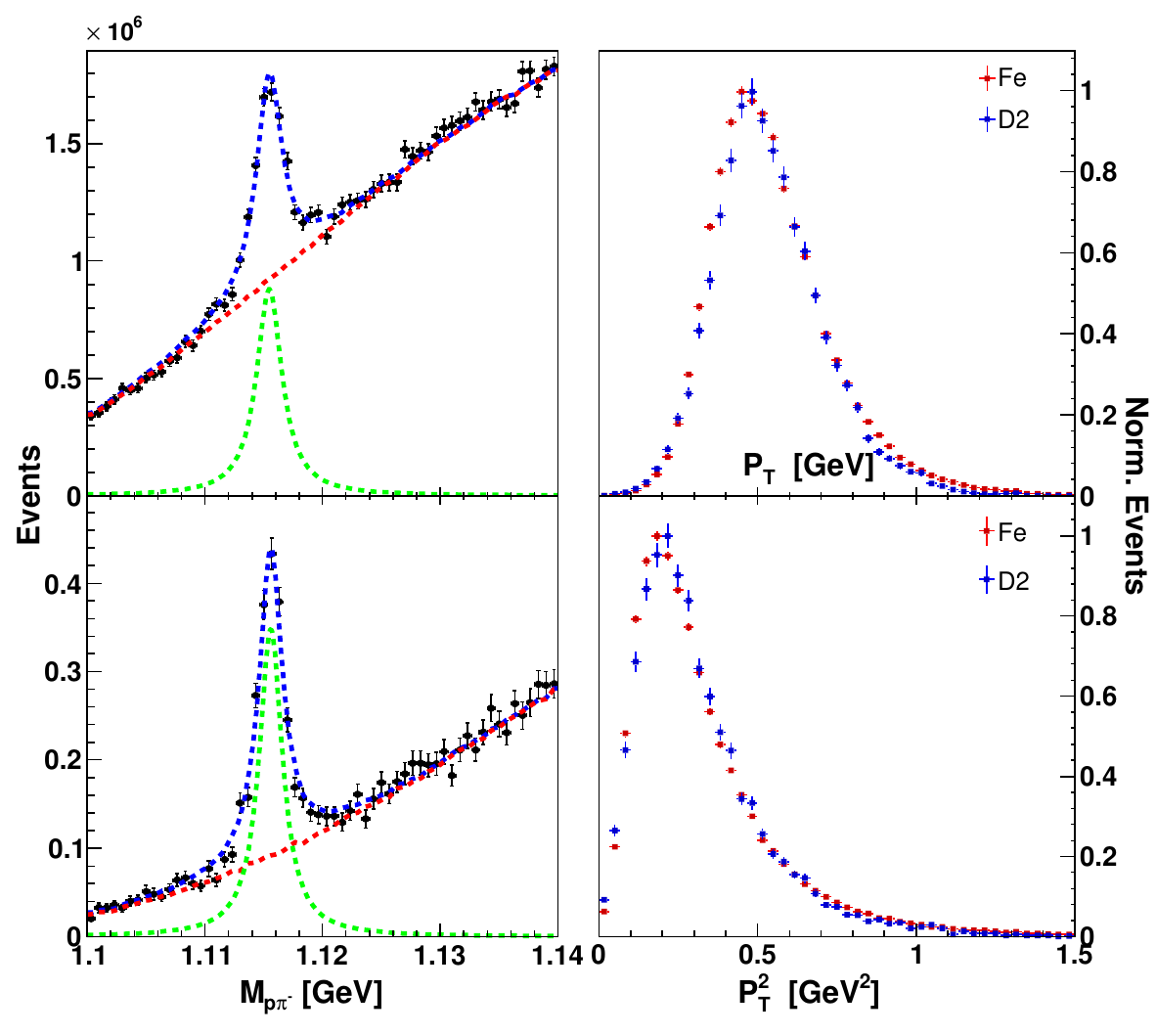}
\vglue -0.2 in
\caption{Left:~acceptance-weighted $(p, \pi^-)$ invariant mass distributions for the~Fe/LD2~(top/bottom)~targets.~Blue curves represent~the~{\tt RooFit}~$\chi^2$ minimization using a simple Breit-Wigner (BW) function for the $\Lambda$ signal and event mixing for the combinatorial background (red dotted curves).~The green distributions are the fit results that are integrated to obtain the $\Lambda$ yields.~Right:~comparison~of~Fe~(red)~and LD2~(blue)~acceptance-weighted $p_T$/$p_T^2$ (top/bottom)~normalized distributions to their peak height.}\label{fig:mass_spect} \end{figure}

The second observable is the transverse momentum broadening, $\Delta p^2_{T}$, defined as
\begin{equation}
 \Delta p^2_{T}=\, \langle p^2_{T}\rangle_A - \langle p^2_{T}\rangle_{D},
 \label{pT-broad}
\end{equation}
where $\langle p^2_T\rangle_A$ is the mean $p_{T}$ squared for a target~$A$~(see Fig.~\ref{fig:mass_spect} bottom right).~This observable carries crucial information about the interaction of the propagating parton with the surrounding color field in the nucleus. Several models correlate the $p_{T}$-broadening with the parton energy loss triggered by the stimulated gluon bremsstrahlung while crossing the medium in the color-neutralization stage~\cite{Baier:1996sk,Brodsky:1992nq}. Based on the perturbative view of the Lund string model, the propagating quark's energy loss is predicted to be at a rate comparable to its string constant on the order of 1~GeV/fm~\cite{Andersson:1979ue, Andersson:1983ia}.~This effect is believed to be the reason behind the observed jet quenching in heavy-ion collisions at the Relativistic Heavy Ion Collider and at the Large Hadron Collider, leading to the suppression of large $p_{T}$ hadron production in nucleus-nucleus compared to proton-proton collisions~\cite{STAR:2005vxr,PHENIX:2004vcz}.

In this Letter, results on SIDIS production of $\Lambda$ hyperons off nuclei, \textit{i.e.}, $e + A\to e'+\Lambda+ X$, are reported, where $A$ is the heavy nuclear target or deuterium, $X$ is the unobserved hadronic system, and $\Lambda$ is identified in the final state through its decay products $\pi^-$ and $p$. The results represent the first-ever measurement of $\Lambda$ multiplicity ratios and $p_{T}$-broadening as a function of $z$ and the atomic mass-number, $A$, for the latter in the current (forward) fragmentation region, in which the struck (di)quark initiates the hadronization process, and the target (backward) fragmentation region, in which the target remnant moves reciprocally with regard to the $\gamma^*$ direction undergoing a spectator or target fragmentation. Furthermore, the current and target fragmentation processes are assumed to have dominant contributions in distinct phase space regions, which are kinematically separated via the coverage of the Feynman scaling variable $x_F$~\cite{Ceccopieri:2012rm, Ceccopieri:2015kya}.

Previous measurements of $R^A_h$ for various hadrons, mainly mesons and (anti)protons by the HERMES \cite{Airapetian:2000ks,Airapetian:2003mi,Airapetian:2007vu,Airapetian:2009jy,Airapetian:2011jp} and the CLAS~\cite{Daniel:2011nq, Moran2022} Collaborations have reported a strong suppression of leading hadrons at high $z$ and a slight enhancement of multiplicity ratios at low $z$ while scanning heavy to light nuclei.~This inverted effect for slow (backward) and fast (forward) protons in HERMES results, the sole baryon study so far, demonstrates the importance of separating the two regions to properly interpret the data.~Approximate separation is possible via the $z$ dependence of the Feynman variable $x_F$~\footnote{The frame-dependent Feynman variable~$x_{F}= \,p_L^*/p_{L^{max}}^*$ is defined as the fraction of the center-of-mass longitudinal momentum carried by the hadron with respect to the $\gamma^*$ direction in the lab frame.} given that the current fragmentation (high~$z$) is dominated by positive~$x_F$, while the target remnant favors negative $x_F$~\cite{Ceccopieri:2012rm, Ceccopieri:2015kya, Graudenz:1994dq}.

A study of $\Delta p^2_{T}$~for mesons was also performed by the HERMES experiment~\cite{Airapetian:2009jy}, but its finding could not distinguish between models predicting an $A^{1/3}$ or $A^{2/3}$ mass dependence \cite{Baier:1996sk,Brodsky:1992nq}.~The $\Delta p^2_{T}$ is expected to increase linearly as $A^{1/3}$ if it is proportional to the nuclear radius and thus the crossed path length, $L$, in the nuclear medium, while an increase as $A^{2/3}$ would indicate a dependence on partonic energy loss via the prediction that $\frac{\Delta E}{dx} \propto \Delta p^2_{T}$ and thus $\Delta E \propto L^2$~\cite{Baier:1996sk}. 

The data presented in this paper were collected during early 2004.~An electron beam of 5.014~GeV energy was incident simultaneously on a 2-cm-long liquid-deuterium target (LD2) and a 3-mm-diameter solid target (carbon, iron, or lead).~A~remotely controlled dual-target system~\cite{Hakobyan:2008zz} was used to reduce systematic uncertainties and allow high-precision measurements of various experimental observables~\cite{Moran2022,ELFASSI2012326}.~The cryogenic and solid targets were located 4~cm apart to minimize the difference in CLAS acceptance, while maintaining the ability to identify event-by-event the target where the interaction occurred via vertex reconstruction~\cite{Schmookler:2019nvf}. The thickness of each solid target (1.72~mm for C, 0.4~mm for Fe, and 0.14~mm for Pb) was chosen so that all targets including deuterium would have comparable per-nucleon luminosities ($\sim$10$^{34}$~cm$^{-2}$s$^{-1}$).~The scattered electrons, negative pions, and protons were detected in coincidence using the CLAS spectrometer~\cite{MECKING2003513}.~The scattered electrons were identified requiring a coincidence between the Cherenkov counter~and the electromagnetic calorimeter~signals~\cite{ELFASSI2012326}, while pions and protons were identified through time-of-flight measurements~\cite{ELFASSI2012326, Schmookler:2019nvf, Hen:2012yva}.

The $\Lambda$ hyperons were identified through the reconstructed invariant mass of detected pions and protons (see the first section of the Supplemental Material (SP.1) for more details about the $\Lambda$ identification method~\cite{Supp}). For each event, several kinematic variables were evaluated including $Q^2$, the virtual photon-nucleon invariant mass squared $W^2$, and the $\gamma^*$ energy fraction $y = \nu/E_e$, where $E_e$ is the incident beam energy. The SIDIS $\Lambda$ events were selected with $Q^2$ > 1~GeV$^2$ to probe the nucleon structure, $W > $ 2~GeV to suppress contamination from the resonance region, and $y <$ 0.85 to reduce the size of radiative effects on the extracted multiplicity ratios based on the HERMES studies~\cite{Airapetian:2000ks,Airapetian:2003mi,Airapetian:2007vu,Airapetian:2009jy,Airapetian:2011jp}. The ($p, \pi^-$) invariant mass distributions are shown in Fig.~\ref{fig:mass_spect} left for iron (top) and LD2 (bottom) with all cuts applied. The distributions exhibit a clean $\Lambda$ peak positioned around 1115.7~MeV sitting on a substantial combinatorial background (CB). An advanced data modeling and fitting toolkit {\tt RooFit}~\cite{ROOFit} was used along with the event mixing technique to subtract the CB (red dotted curves in Fig.~\ref{fig:mass_spect} left), which is reconstructed by combining uncorrelated $p$ and $\pi^-$ tracks from different events~\cite{TM-thesis}.~The extraction of the background-subtracted $\Lambda$ yields, as well as the $p^2_{T}$ means, was performed after weighting their distributions event-by-event with the inverse of the acceptance correction (AC) factors.~The latter were evaluated using events generated with the Pythia event generator~\cite{Gallmeister:2005ad} and processed by the CLAS GEANT3 package~\cite{GSIM} to simulate the detector geometrical acceptance, as well as the associated detection and reconstruction efficiencies.~Pythia was modified to include nuclear parton distribution functions~\cite{Buckley:2014ana} and Fermi motion based on the Paris potential distribution and realistic many-body calculations~\cite{CiofidegliAtti:1995qe}.~Radiative effects were also included in the simulation using the RadGen code~\cite{Akushevich:1998ft} developed to correct lepton-nucleon scattering observables from quantum electrodynamics radiative processes.~Small corrections were also applied for other effects related to proton energy loss, scattering angle and momentum distortions, vertex misalignment~\cite{Schmookler:2019nvf, Hen:2012yva}, and LD2 end-cap contamination.
\begin{figure*}[t]
\centering
\includegraphics[clip=true, trim= 0.95cm 0.35cm 7cm 5.5cm,width=0.495\textwidth, keepaspectratio=TRUE]{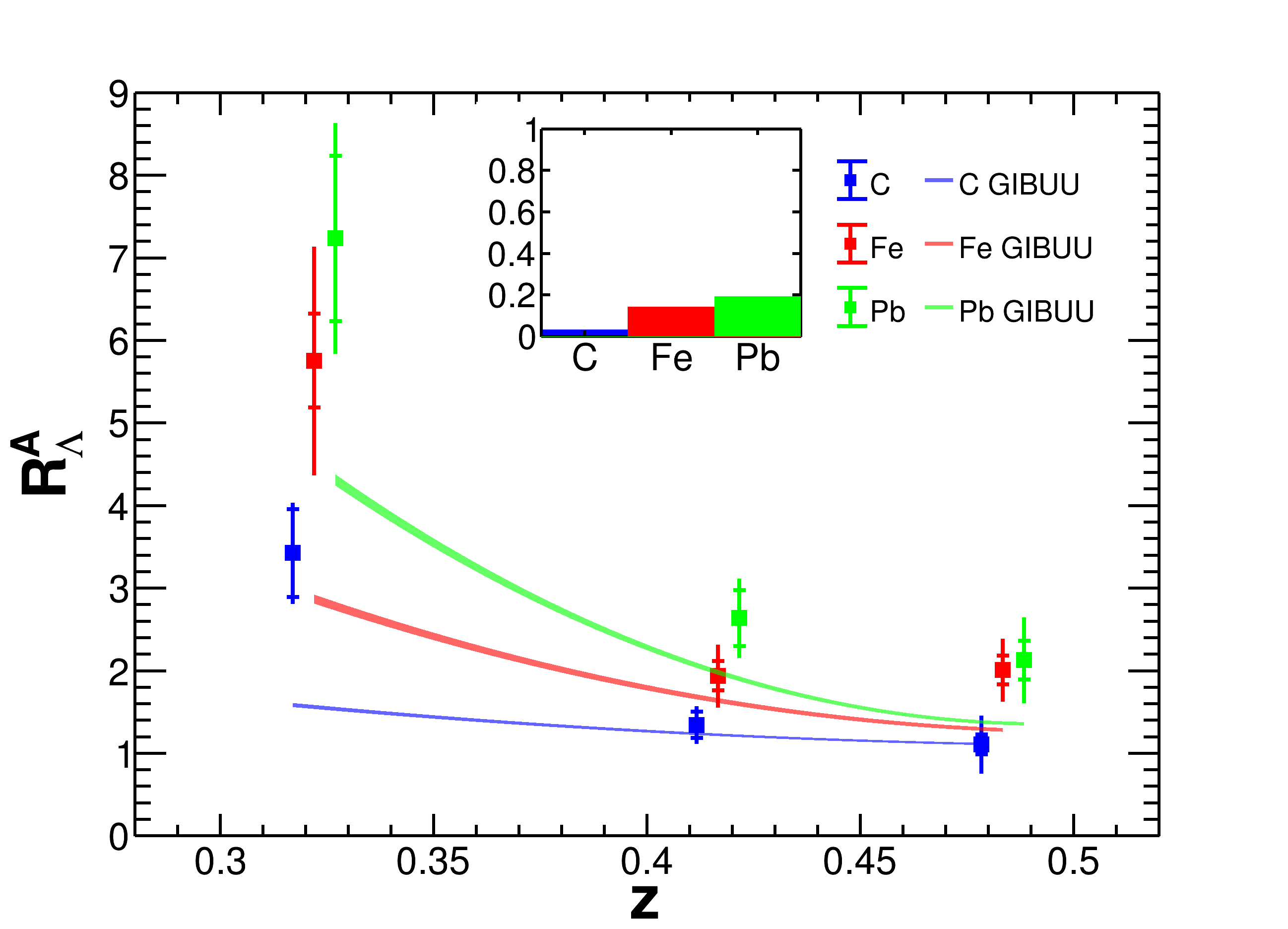}
\includegraphics[clip=true, trim= 0.9cm 0.35cm 8cm 0,width=0.495\textwidth,keepaspectratio=TRUE]{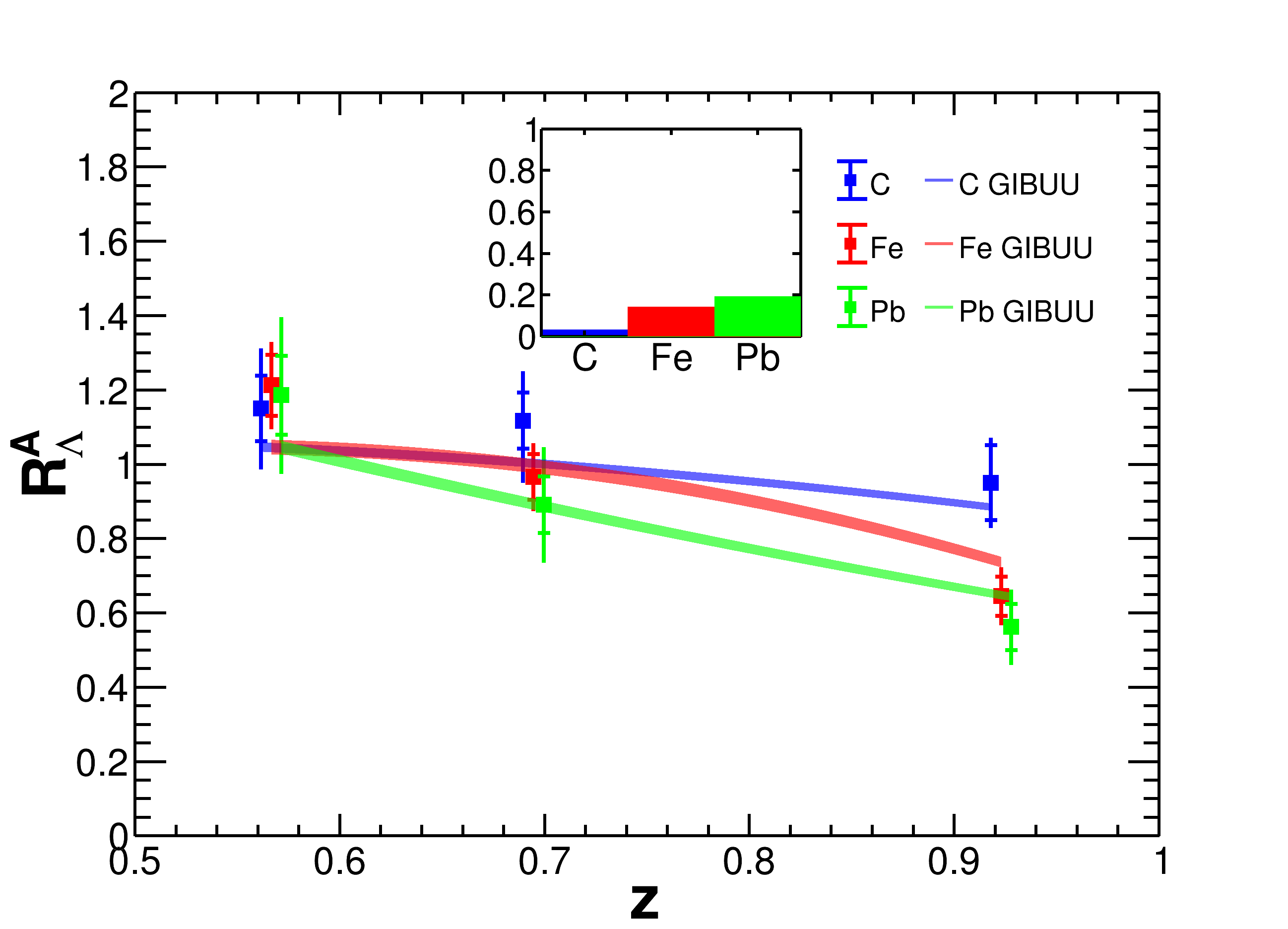}
\vglue -0.145 in
\caption{$\Lambda$~$z$-binned multiplicity ratios for carbon, iron, and lead (the results are horizontally shifted for clarity). The outer error bars are the $p2p$ systematic uncertainties added in quadrature with the statistical uncertainties.~The inset contains the total normalization uncertainties for each nucleus.~The plots illustrate the results of the low~(left)~and~high~(right)~$z$~ranges corresponding, respectively, to the target and current fragmentation regions.~The curves correspond to GiBUU model calculations~\cite{Buss:2011mx}.} 
\label{Fig:mratioWithSystematics_shifted}
\end{figure*}

\begin{figure*}[th!]
\includegraphics[clip=true, trim= 0.3cm 0.35cm 7cm 5.75cm,width=0.495\textwidth,keepaspectratio=TRUE]{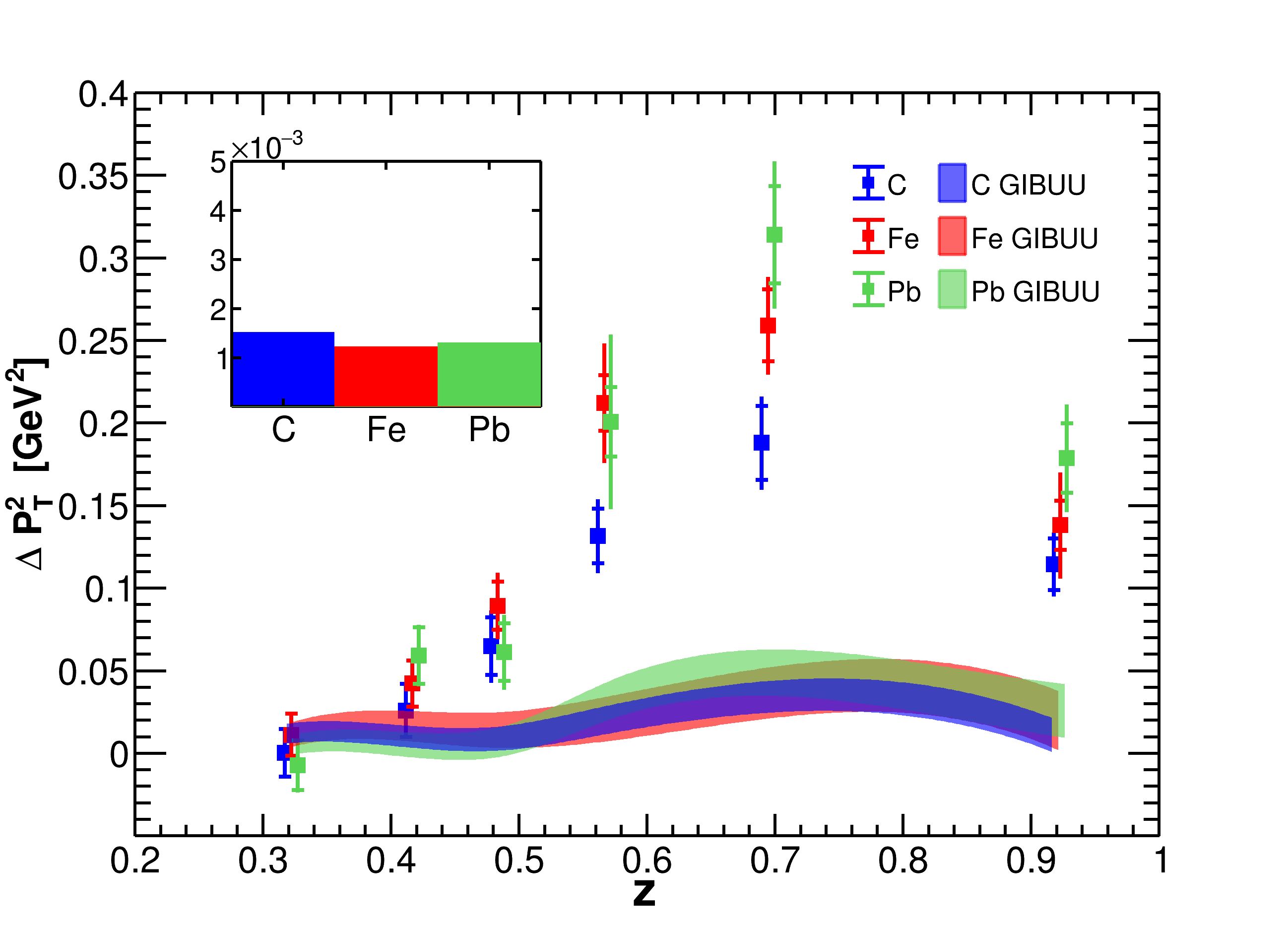}
\includegraphics[clip=true, trim= 3.8cm 0.35cm 7cm 3.5cm,width=0.48\textwidth, keepaspectratio=TRUE]{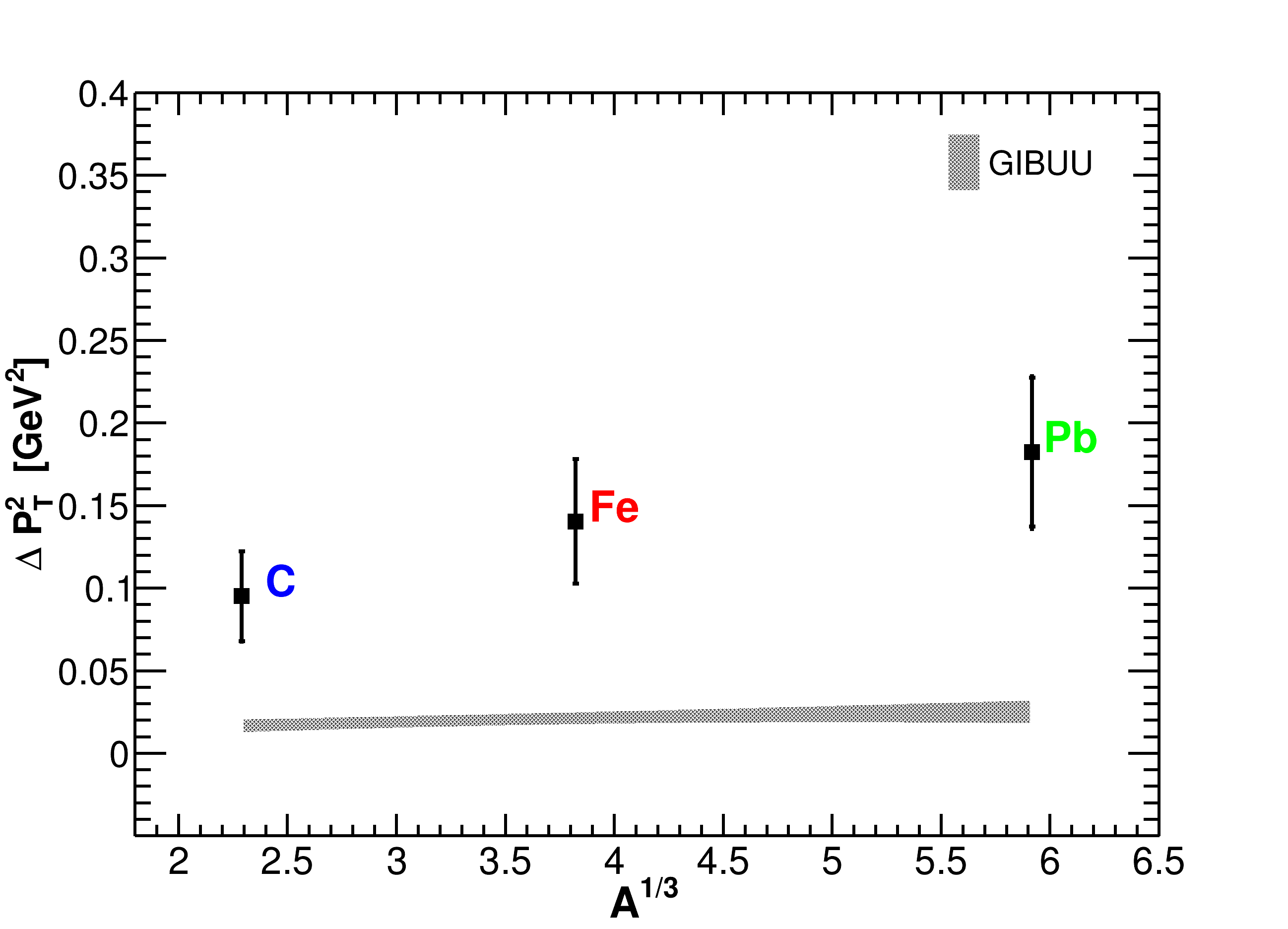}
\vglue -0.15in \caption{Left (right): the $z$ (nuclear radius)-dependent $\Delta p^2_T$ results for the three nuclei (results are horizontally shifted for clarity). The outer error bars are the $p2p$ systematic uncertainties added in quadrature with the statistical uncertainties, while the normalization uncertainties are presented in the inset for the $z$ dependence and found to be less than 1\% for the $A$ dependence. The GiBUU model calculations are represented by the colored (left) and shaded (right) bands obtained by interpolating the model points and their statistical uncertainties.}
\label{Fig:pT2GIBUU}
\end{figure*}

Due to the limited statistics of the $\Lambda$ production channel, the extractions of both multiplicity ratios and $p_{T}$-broadening results were performed by integrating over all kinematic variables except $z$,~which is divided into the six bins shown in Table~\ref{tab:zBinning} of the Supplemental Material~\cite{Supp}.~Given that the interest in this work is in the $z$ and $A$ dependencies of the observables, the systematic uncertainties were separated into point-to-point ($p2p$), which exhibit some $z$ and $A$ dependencies, and normalization uncertainties, which are kinematics independent.~An in-depth study was carried out and the main systematic sources are related to~1) particle identification cuts to identify the three final-state particles, scattered electron, $p$, and $\pi^-$,~2) dual-target vertex corrections,~3) AC multidimensional (6D) map variables and the binning that was chosen based on the comparison of experimental data and simulation,~4) AC weight cuts to suppress artificial spikes due to poor statistics in some AC 6D bins,~5)~CB subtraction methods by varying the event mixing uncorrelated track combinations and BW shapes utilized in {\tt RooFit} for $R^A_{\Lambda}$ while considering CB sideband subtraction for $\Delta p^2_T$,~6)~$\Lambda$ mass range for $R^A_{\Lambda}$,~and~7) LD2 end caps and radiative correction procedures.~As a result, the total p2p (normalization) uncertainties vary between 6\%~to~30\% (less than 3\%) for the multiplicity ratios of all nuclei with the dominant contributions from the AC and CB subtraction methods (see Table~\ref{mrSysTable}~\cite{Supp}).~Similarly, the total p2p uncertainties vary between 10\%~(1.4\%) and~81\%~(8.5\%) for the nuclear $z$~($A$)~dependence of $p_{T}$-broadening (see Table~\ref{pTBroadZ-SysTable}~(\ref{pTBroadA-SysTable})~\cite{Supp}),~while the total normalization uncertainty for both dependencies is less than 1\%.~The largest p2p $z$-dependent uncertainty, which is associated with the lead target, is still less than the~50\% statistical uncertainty as shown in Fig.~\ref{Fig:pT2GIBUU}.

The $\Lambda$ multiplicity ratio results are depicted in Fig.~\ref{Fig:mratioWithSystematics_shifted} along~with~theoretical calculations~from~the Giessen Boltzmann-Uehling-Uhlenbeck (GiBUU) model~\cite{Buss:2011mx}.~As expected, $R^A_\Lambda$ manifests an inverted behavior in the two $z$ regions; at high $z$ (see Fig.~\ref{Fig:mratioWithSystematics_shifted} right), the region in which the current fragmentation dominates, $\Lambda$ baryons exhibit less attenuation in lighter nuclei and greater suppression with $z$, up to 40\% in lead and 35\% in iron at the highest $z$ bin. However, at low $z$ (see Fig.~\ref{Fig:mratioWithSystematics_shifted} left) $R^A_\Lambda$ is more enhanced on heavy nuclei as a signature of the significant contribution from the target fragmentation that predominates in this kinematic region. This observation is consistent with the fact that the $\Lambda$ baryons show a significant leading particle effect; \textit{i.e.}, they carry a substantial fraction of the incoming proton momentum~\footnote{By convention of the $\gamma^* p$ frame, a negative longitudinal momentum fraction $x_{F}$ corresponds to final state hadrons moving parallel to the incoming proton direction, thus covering the low $z$ region. However, the positive $x_{F}$ favors high $z$, where forward fragmentation governs.} and thus large negative $x_{F}$ (see Fig.~\ref{fig:zvsxf}~\cite{Supp}) and small $p_{T}$ relative to the $\gamma^*$ direction~\cite{Ceccopieri:2012rm, Ceccopieri:2015kya}.~The data are qualitatively described by GiBUU for most of the $z$ range and most of the targets except for the lowest $z$ bin, where approximately a factor of two difference is observed.

Figure~\ref{Fig:pT2GIBUU} contains the $\Lambda$ $p_{T}$-broadening results as a function of~$z$~(left)~and~$A$~(right)~along with theoretical calculations from the GiBUU model~\cite{Buss:2011mx}.~The monotonic increase of broadening with $z$ and the mass-number reflects the interaction of the propagating object with the surrounding color field in the nucleus during the neutralization stage and/or the elastic scattering of the prehadron and the fully formed $\Lambda$~\cite{Baier:1996sk,Brodsky:1992nq}.~Such a (pre)hadron interaction, as well as broadening, seems to diminish at the highest $z$ bin.~This is an indication of the partonic stage dominance of the hadronization process preceding the (pre)hadron formation, as their elastic scattering in the medium should have led to more broadening as $z$ approaches unity~\cite{Airapetian:2009jy, DiNezza:2008zz}.~This trend is in favor of the $A^{1/3}$ dependence of $\Delta p_T^2$ and implies that the production time is within the nuclear medium.~Yet, the measured $\Lambda$ hyperon broadening is an order of magnitude greater than that seen in the HERMES meson results~\cite{Airapetian:2009jy}. This could be due to the quark-diquark nucleon structure so that the virtual photon, instead of being absorbed by a quark, is absorbed by a diquark. That is to say, the propagating colored diquark has a sizable mass and an extended QCD color field compared to a single quark, leading to more in-medium interactions, and thus an increase of the $\Delta p^2_T$ magnitude~\cite{diquark2021}.~This diquark scattering speculation offers a good explanation of the $R^A_\Lambda$ attenuation with increasing $z$ in the current fragmentation region. While GiBUU has reasonably described HERMES, EMC~\cite{Gallmeister:2007an, Falter:2003di, Falter:2004uc}, and CLAS~\cite{Daniel:2011nq, Moran2022} multiplicity ratio measurements, it underestimates our $\Lambda$ $p_{T}$-broadening results, which could indicate that the angular distribution is inaccurate in the initial elementary production process of $\Lambda$ or that the final state interactions in the current model's string fragmentation functions are not realistic~\cite{GallMosel22}.

In summary, the first-ever measurement of $\Lambda$ multiplicity ratios and $p_{T}$-broadening as a function of $z$ and $A$ in the current and target fragmentation regions are reported. Both observables depend strongly on $z$, with an enhancement of $R^A_\Lambda$ at low $z$ and a suppression at high $z$ up to 0.951 $\pm$ 0.125 for carbon, 0.645 $\pm$ 0.164 for iron, and 0.562 $\pm$ 0.219~\footnote{The quoted uncertainty values of 0.125 for carbon, 0.164 for iron, and 0.219 for lead targets are simply the quadrature sum of the statistical and $p2p$ systematical uncertainties shown in Table~\ref{tab:mr} for the highest $z$ bin~\cite{Supp}.} for lead, and an increase of $p_{T}$-broadening with $A$ and $z$ except for the last $z$ bin where the broadening starts decreasing due to the partonic stage dominance of the hadronization process. The one order of magnitude larger broadening for this hyperon channel compared to HERMES meson results, as well as the strong suppression of $R^A_\Lambda$ at high $z$, suggests the possibility of a direct scattering off diquark configurations of the nucleon. The multiplicity ratio results are qualitatively described by the GiBUU transport model, however, the model strongly underestimates our $p_{T}$-broadening results. This finding has the potential to stimulate further experimental and theoretical investigations, constrain existing models such as GiBUU, and open a new era of studies of nucleon and light hyperon structure.

Future higher-luminosity measurements with CLAS12 and an 11~GeV~beam energy~\cite{BrooksProposal} will study SIDIS production of a variety of mesons and baryons over a wide kinematic range.~This is crucial to constrain competing models and boost our understanding of the fragmentation mechanisms that lead to the formation of various hadrons. It would also provide an opportunity to study for the $\Lambda$ SIDIS final states the correlation between kaons and $\Lambda$'s that will presumably be sensitive to the diquark structure in the struck nucleon. The forthcoming experiments with CLAS12, in addition to measurements at the planned Electron Ion Collider~\cite{Accardi:2012qut}, have the potential to investigate in great detail the speculated diquark scattering in the current results, which would have a significant impact on our understanding of nucleon and baryon structure.
\section*{\label{acknowledgement}\bf{Acknowledgments}}

The authors would like to thank K.~Gallmeister and U.~Mosel for the fruitful discussions on the GiBUU model predictions for this $\Lambda$ production channel. We acknowledge the staff of the Accelerator and Physics Divisions at the Thomas Jefferson National Accelerator Facility who made this experiment possible.~This work was supported in part by the U.S. Department of Energy Award~No.~DE-FG02-07ER41528, the Physics and Astronomy Department and the Office of Research and Economic Development at Mississippi State University, the Chilean Agencia Nacional de Investigacion~y~Desarrollo~(ANID), including by~ANID~PIA/APOYO~AFB180002,~ANID PIA ACT1413, and~ANID – Millennium~Program – ICN2019\_044,~the U.S.~Department~of~Energy,~Office of Nuclear Physics, under Contract~No.~DE-AC02-06CH11357, by the Italian Istituto Nazionale di Fisica Nucleare, the French Centre National de la Recherche Scientifique, the French Commissariat á l’Energie Atomique, the United Kingdom Science and Technology Facilities Council (STFC), the Scottish Universities Physics Alliance (SUPA), the National Research Foundation of Korea, and the U.S. National Science Foundation.~The Southeastern Universities Research Association operates the Thomas Jefferson National Accelerator Facility for the U.S.~Department of Energy under Contract No.~DE-AC05-06OR23177.

\bibliography{bibio} 
\clearpage

\newpage
\pagebreak
\onecolumngrid
\appendix*
\section*{\large Supplemental Material}
\setcounter{equation}{0}
\setcounter{figure}{0}
\setcounter{table}{0}
\setcounter{page}{1}
\renewcommand{\theequation}{S\arabic{equation}}
\renewcommand{\thefigure}{S\arabic{figure}}
\renewcommand{\thetable}{S\arabic{table}}
\renewcommand{\bibnumfmt}[1]{[S#1]}
\renewcommand{\citenumfont}[1]{S#1}
\renewcommand{\tabcolsep}{6pt}
\renewcommand{\arraystretch}{1.5}
\newcounter{SIfig}
\renewcommand{\theSIfig}{S\arabic{SIfig}}

\begin{figure}[phb!]
\refstepcounter{SIfig}\label{fig:zvsxf}
\includegraphics[width=0.48\textwidth,keepaspectratio=TRUE]{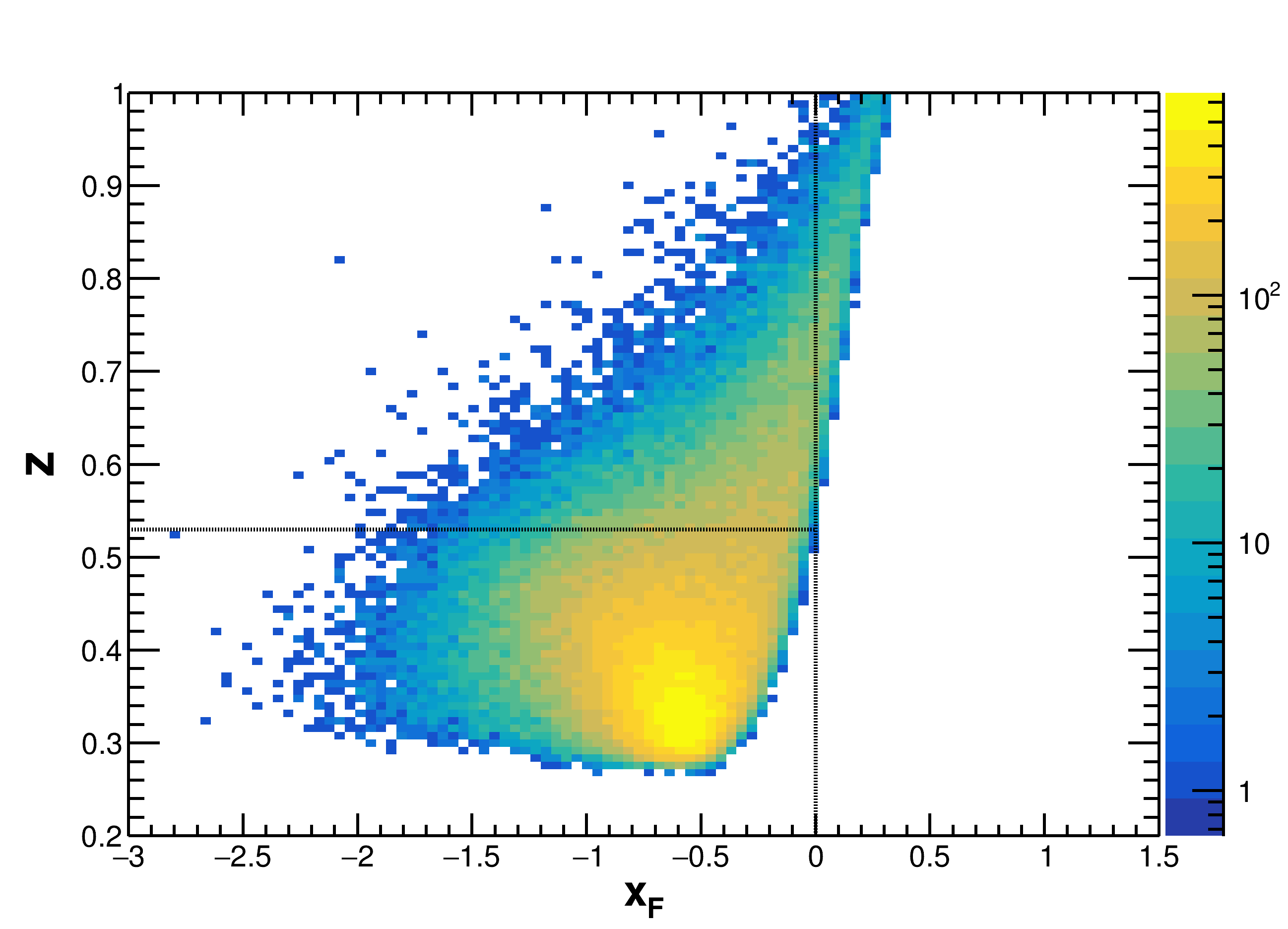}
\vglue-0.25cm\caption{$z$~vs.~$x_{F}$, where the horizontal dashed line around values of $z$ greater than $\sim$0.55 depicts the discussed separation between forward and backward fragmentation regions suggested by the sign change of $x_{F}$ (vertical dashed line).}
\end{figure}

\begin{turnpage}
\begin{figure}[b]
\centering
\includegraphics[clip=true, trim= 1.95cm 0.02cm 0.95cm 0.5cm,width=0.47\textwidth, keepaspectratio=TRUE]{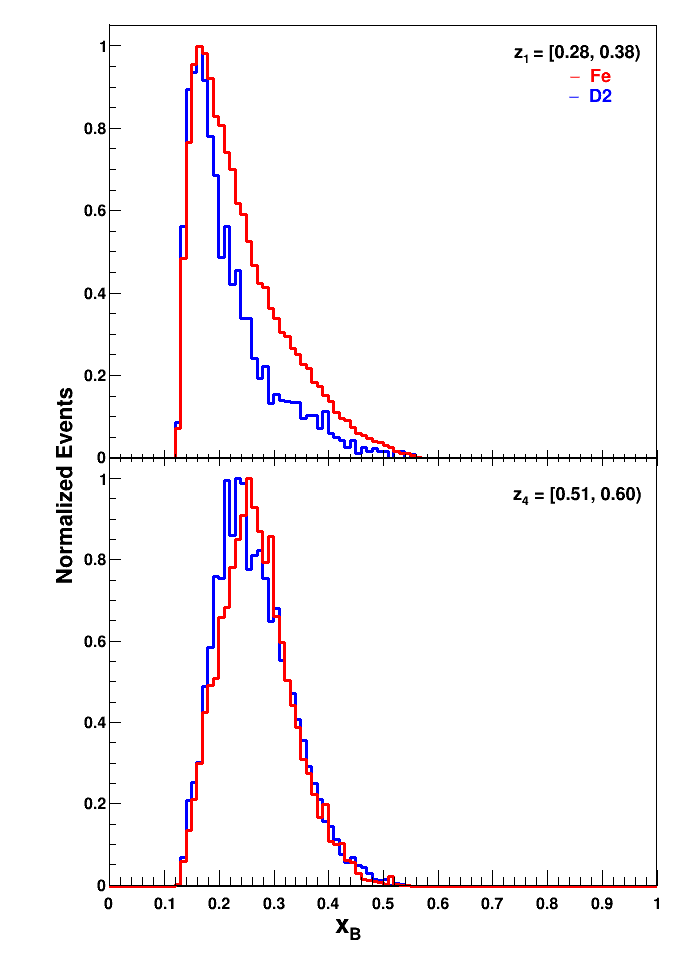}\includegraphics[clip=true, trim= 3cm 0.02cm 0.95cm 0,width=0.445\textwidth, keepaspectratio=TRUE]{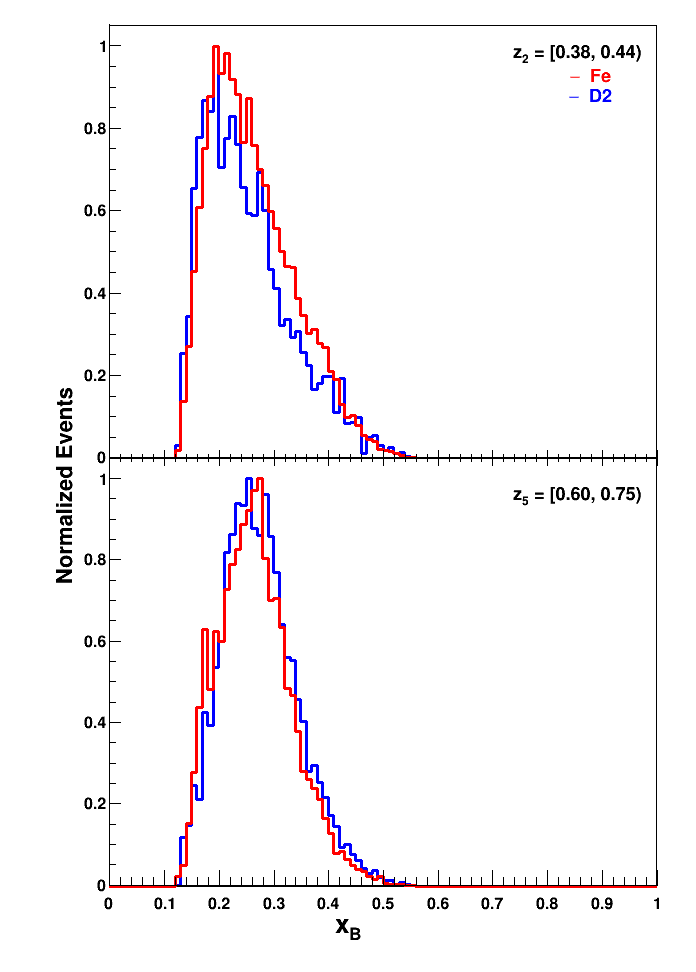}\includegraphics[clip=true, trim= 3cm 0.02cm 0.95cm 0,width=0.445\textwidth, keepaspectratio=TRUE]{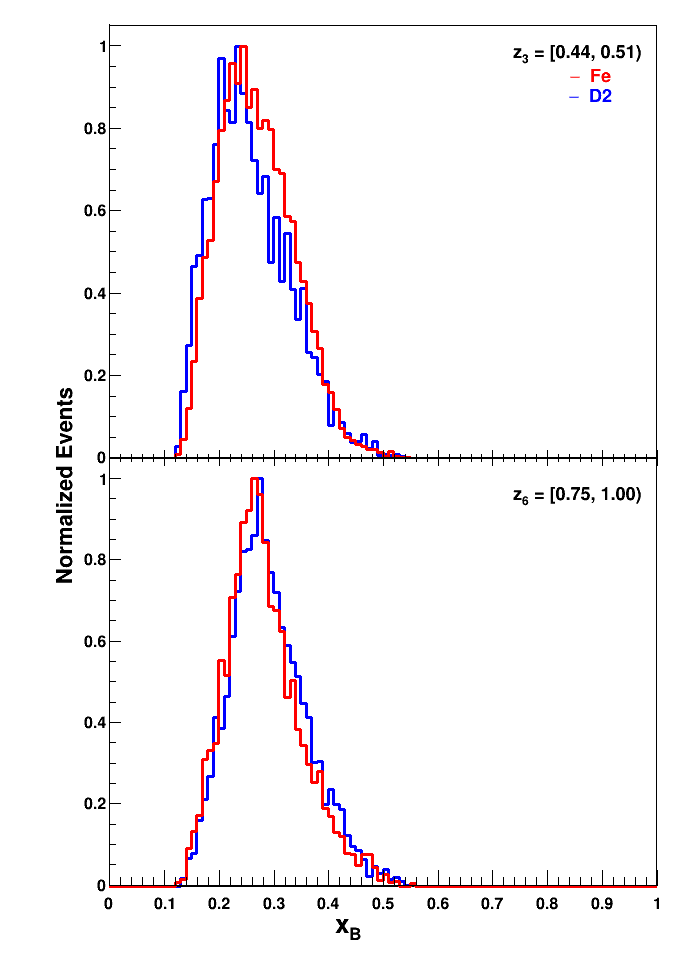}
\vglue -0.15 in
\caption{Comparison~of the $z$-binned~Fe~(red)~and LD2~(blue)~normalized acceptance-weighted $x_B$ distributions for the depicted $z$ bins that are defined in Table~\ref{tab:zBinning}.}\label{fig:xB} \end{figure}
\end{turnpage}

This appendix contains supplementary information about the $\Lambda$ identification method in~{\bf SP.1}, the acceptance correction details related to the multidimensional (6D) map variables and binning, weight definition and cut, and its application procedure in {\bf SP.2}, a summary of the contributions of systematic effects to the total point-to-point uncertainty budget in {\bf SP.3}, and the reported results in the last two figures of this manuscript, Figs.~\ref{Fig:mratioWithSystematics_shifted} and~\ref{Fig:pT2GIBUU}, as well as two supporting figures, Figs.~\ref{fig:zvsxf} and~\ref{fig:xB}, in {\bf SP.4}. In Table~\ref{tab:mr}, the $z$-binned multiplicity ratios are given for all nuclei, while Table~\ref{tab:pt2z} (Table~\ref{tab:pt2A}) contains the transverse momentum broadening as a function of $z$ ($A$) for all nuclei. \\

\paragraph{\bf{SP.1~Lambda Identification}}
\label{LambdaID}
In the sample of reconstructed SIDIS events originating from either the liquid or solid target, one scattered $e^-$ and at least one $\pi^-$ and $p$, the decay products of the $\Lambda$, were required. To reconstruct the $z$ binned ($\pi^-$, $p$) invariant mass spectrum for each target, the 4-vector energy-momentum ($P^\mu= (E, p_x, p_y, p_z)$) of all identified negatively charged pions and protons were combined event-by-event as
\begin{equation}
P_\Lambda = P_p + P_{\pi^{-}},
\label{InvMassEq}
\end{equation}
where $P_{\Lambda}$, $P_p$, and $P_{\pi^{-}}$ are the 4-vector energy-momentum of the $\Lambda$ candidates, protons, and $\pi^-$s, respectively. Figure~\ref{fig:mass_spect} left shows the acceptance-weighted invariant mass from solid (top) and liquid (bottom) targets in which the $\Lambda$ peak sits on a huge combinatorial background (red dotted curves) that is subtracted using {\tt RooFit} to extract the pure $\Lambda$ yields and thus obtain the presented multiplicity ratios in Fig.~\ref{Fig:mratioWithSystematics_shifted}.\\
\paragraph{\bf{SP.2~Acceptance Correction}}
\label{LambdaACC}
The adopted acceptance correction for this analysis is based on a bin-by-bin correction method. Its main advantage is that it should be, in principle, independent of the model used in the  Monte-Carlo (MC) event generator if the chosen bins are infinitely small. This is very important for this channel since it is not expected that the employed model in Pythia would be realistic enough to perfectly reproduce the data. Based on a comparison between MC and experimental data, the chosen AC six dimensional (6D) map variables and binning are summarized in Tables~\ref{tab:accCorrBinMap}-~\ref{tab:zBinning}.
\begin{table}[h!]
\centering
\begin{tabular}{|c|c|c|c|}
\hline
 Variables                & Range      & Number of bins & Bin width                  \\ \hline \hline
 $W$ [GeV]                & 2.00 - 2.80    &       2        & 0.4                      \\ \hline
 $\nu$                    & 2.25 - 4.25    &       3        & 0.$\overline{6}$         \\ \hline
 $\phi_{\pi^{-}}$ [deg]   & 0.0 - 360.0    &       2        & 180                      \\ \hline
 $\phi_{e\Lambda}$ [deg]& 0.0 - 360.0    &       3        & 120                      \\ \hline
 $P_{\Lambda}$ [GeV]  & 0.10 - 4.25    &       3        & 1.38$\overline{3}$       \\ \hline
 $z$                      & 0.28 - 1.00    &       6        & {\normalsize see} Table~\ref{tab:zBinning} \\ \hline \hline
 Total                    &              &      648       &                          \\ \hline
\end{tabular}
\caption{Binning for the AC map, where $\nu$, $W$, and $z$ were already defined, $\phi_{\pi^-}$ is the $\pi^-$ azimuthal decay angle in the $\Lambda$ rest frame, $\phi_{e\Lambda}$ is the angle between the leptonic and hadronic planes, and $p_{\Lambda}$ is the $\Lambda$ momentum. Table~\ref{tab:zBinning} shows the $z$ bins used as reported in Table~\ref{tab:mr}.}
\label{tab:accCorrBinMap}
\end{table}

\begin{table}[h!]
\centering
\begin{tabular}{|c|c|c|c|c|c|c|}
\hline
 $z-$bin \# & 1 & 2 & 3 & 4 & 5 & 6  \\ \hline
 $z_{min}$ & 0.28 & 0.38 & 0.44 & 0.51 & 0.60 & 0.75   \\ \hline
 $z_{max}$ & 0.38 & 0.44 & 0.51 & 0.60 & 0.75 & 1.00   \\ \hline
\end{tabular}%
\caption{The $z$ bins used in this analysis.}
\label{tab:zBinning}
\end{table}

The acceptance correction factors are defined for each 6D bin 
$k$= ($W$, $\nu$, $p_{\Lambda}$, $\phi_{\pi^-}$, $\phi_{e\Lambda}$, $z$) as 

\begin{equation}
\label{eq:effk}
AC_k= \displaystyle{\frac{N_{acc}(W, \nu, p_{\Lambda}, \phi_{\pi^-}, \phi_{e\Lambda}, z)}{N_{gen}(W, \nu, p_{\Lambda}, \phi_{\pi^-}, \phi_{e\Lambda}, z)}},
\end{equation}

\noindent where $N_{gen}(W, \nu, p_{\Lambda}, \phi_{\pi^-}, \phi_{e\Lambda}, z)$ and $N_{acc}(W, \nu, p_{\Lambda}, \phi_{\pi^-}, \phi_{e\Lambda}, z)$ are, respectively, the number of generated and accepted events in each bin $k$. Once these {AC} coefficients were computed, the data were corrected event-by-event by a weight $\omega_k = 1/AC_k$, which depends on the bin $k$ to which it belongs. It should be noted that if some 6D AC bins have very small correction factors due to their poor statistics, an artificially large weight would be attributed to those bins that would lead to spikes in the weighted distributions. To avoid this problem, the following weight cut was adopted to minimize this effect on the weighted distributions: 
\begin{eqnarray}
\label{eq:cut-w}
&& 60 < \omega_k \leq 2400. \\ \nonumber
\end{eqnarray}

Furthermore, the effect of this weight cut was estimated and applied as a global correction factor, $f_\omega$, to the extracted results. This estimation was done by weighting the MC accepted $N_{acc}$ events and comparing their sum, $\sum \omega N_{acc}$, to the generated ones as
\begin{equation}
f_\omega = \frac{\sum \omega N_{acc}}{N_{gen}.}
\label{fweight}
\end{equation}
\noindent
This $N_{acc}$ weighted sum is typically equal to the generated events without the weight cut, however, it is slightly less once applied, leading to various $f_\omega$ corrections for each $z$-binned multiplicity ratio result as the $p_T$-broadening means are insensitive to this correction.\\

\paragraph{\bf{SP.3~Systematic Uncertainties Budget}}
\label{SysBud}

This section contains the contribution of various systematic effects to the reported total point-to-point systematic uncertainty budget for the $\Lambda$ multiplicity ratios of all nuclei in Table~\ref{mrSysTable} and the corresponding $z$~($A$)~dependence of $p_{T}$-broadening in Table~\ref{pTBroadZ-SysTable}~(\ref{pTBroadA-SysTable}).\\ 
\begin{turnpage}
\begin{table}[bh!]
\Large
\centering
\caption{Multiplicity ratio systematic effects and their contributions for the $z$ bins shown in Table~\ref{tab:zBinning}.}
\label{mrSysTable}
\resizebox{\textheight}{!}{%
\begin{tabular}{|c|c|c|c|c|c|c|c|c|c|c|c|c|c|c|c|c|c|c|}
\hline
 \multirow{3}{*}{Systematic Effect} &
  \multicolumn{18}{c|}{$z$ bin Point-to-point Systematic Uncertainty ($\%$)} \\\cline{2-19} 
  &
  \multicolumn{6}{c|}{Carbon}&
  \multicolumn{6}{c|}{Iron} &
  \multicolumn{6}{c|}{Lead} \\ 
  \cline{2-19}
 &
$z-$1 & $z-$2 & $z-$3 &  $z-$4 & $z-$5 & $z-$6 &  
$z-$1 & $z-$2 & $z-$3 &  $z-$4 & $z-$5 & $z-$6 & 
$z-$1 & $z-$2 & $z-$3 &  $z-$4 & $z-$5 & $z-$6   \\ \hline \hline
Particle identification cuts   &   	0.69 	&	4.24 	&	7.24 	&	1.53 	&	3.16 	&	0.00 	&	
0.00 	&	0.95 	&	4.34 	&	0.87 	&	3.17 	&	4.45 	&	
8.05 	&	3.21 	&	7.80 	&	0.00 	&	8.59 	&	6.91	
\\ \hline
Vertex corrections &   	0.28 	&	0.00 	&	0.04 	&	0.22 	&	0.22 	&	0.54 	&	
1.04 	&	1.28 	&	0.56 	&	0.08 	&	0.00 	&	0.13 	&	
1.38 	&	1.85 	&	0.13 	&	0.18 	&	0.00 	&	1.01	
\\ \hline
AC 6D map variables \& binning   &  	3.28 	&	0.00 	&	6.69 	&	9.97 	&	9.17 	&	2.33 	&	
6.83 	&	4.80 	&	0.00 	&	6.42 	&	5.90 	&	4.93 	&	
6.84 	&	0.00 	&	9.05 	&	7.90 	&	6.06 	&	7.23	
\\ \hline
AC weight cuts  &   	0.00 	&	0.00 	&	10.70 	&	0.70 	&	0.00 	&	0.00 	&	
0.00 	&	0.00 	&	9.17 	&	1.86 	&	0.00 	&	0.00 	&	
5.17 	&	0.00 	&	8.64 	&	12.16 	&	0.00 	&	0.00	
\\ \hline
CB uncorrelated-tracks combinations &   	1.80 	&	0.16 	&	0.37 	&	0.27 	&	0.53 	&	0.00 	&	
1.14 	&	0.14 	&	0.20 	&	0.00 	&	0.36 	&	0.23 	&	
1.79 	&	2.04 	&	0.96 	&	0.13 	&	0.00 	&	0.28	
\\ \hline
Breit-Weigner shapes &   	7.55 	&	10.80 	&	25.75 	&	5.13 	&	8.69 	&	5.77 	&	
20.54 	&	16.37 	&	13.40 	&	1.26 	&	0.46 	&	5.27 	&	
5.77	&	12.02 	&	15.71 	&	4.92 	&	10.85 	&	9.52	
\\ \hline
 $\Lambda$ mass-range  &   	2.10 	&	1.11 	&	0.00 	&	0.86 	&	1.87 	&	2.89 	&	
 2.52 	&	1.52 	&	0.43 	&	0.00 	&	1.35 	&	2.39 	&	
 2.24 	&	1.24 	&	0.00 	&	0.65 	&	1.69 	&	2.72	
 \\ \hline
 LD2 endcaps   &   	0.06 	&	0.00 	&	0.06 	&	0.09 	&	0.11 	&	0.13 	&
 0.03 	&	0.00 	&	0.06 	&	0.08 	&	0.10 	&	0.12 	&	
 0.07 	&	0.00 	&	0.05 	&	0.07 	&	0.09 	&	0.13	
 \\ \hline
Radiative correction  &   	0.00 	&	2.08 	&	1.26 	&	3.18 	&	1.53 	&	0.94 	&	
1.30 	&	1.14 	&	0.29 	&	0.00 	&	0.95 	&	0.21 	&	
0.13 	&	0.00 	&	0.81 	&	1.12 	&	0.58 	&	1.90	
\\ \hline \hline
Total                       &   	8.71 	&	11.84 	&	29.61 	&	11.81 	&	13.25 	&	6.93 	&	
21.86 	&	17.19 	&	16.82 	&	6.86 	&	6.92 	&	8.81 	&	
13.41 	&	12.67 	&	21.58 	&	15.36 	&	15.21 	&	14.20	
\\ \hline
\end{tabular}%
}
\end{table}

\begin{table}[bh!]
\Large
\centering
\caption{Transverse momentum broadening systematic effects and their contributions for the $z$ bins shown in Table~\ref{tab:zBinning}.}
\label{pTBroadZ-SysTable}
\resizebox{\textheight}{!}{%
\begin{tabular}{|c|c|c|c|c|c|c|c|c|c|c|c|c|c|c|c|c|c|c|}
\hline
 \multirow{3}{*}{Systematic Effect} &
  \multicolumn{18}{c|}{$z$ bin Point-to-point Systematic Uncertainty ($\%$)} \\\cline{2-19} 
  &
  \multicolumn{6}{c|}{Carbon}&
  \multicolumn{6}{c|}{Iron} &
  \multicolumn{6}{c|}{Lead} \\ 
  \cline{2-19}
 &
$z-$1 & $z-$2 & $z-$3 &  $z-$4 & $z-$5 & $z-$6 &   
$z-$1 & $z-$2 & $z-$3 &  $z-$4 & $z-$5 & $z-$6 &     
$z-$1 & $z-$2 & $z-$3 &  $z-$4 & $z-$5 & $z-$6   \\ \hline \hline
  Particle identification cuts 	 & 7.14 	&	0.00 	&	3.77 	&	1.77 	&	0.47 	&	6.03 	&	
  1.05 	&	8.19 	&	4.97 	&	0.00 	&	0.82 	&	0.87 	&	
  5.07 	&	2.84 	&	6.24 	&	0.00 	&	3.52 	&	3.24 	
  \\ \hline
  Vertex Corrections	 & 6.63 	&	8.99 	&	4.62 	&	1.02 	&	0.00 	&	3.99 	&	
  2.57 	&	2.54 	&	0.00 	&	0.33 	&	0.33 	&	0.67 	&	
  3.40 	&	1.87 	&	0.81 	&	0.00 	&	0.74 	&	2.79 	
  \\ \hline
  AC 6D map variables \& binning	 & 4.77 	&	6.95 	&	6.02 	&	0.00 	&	8.91 	&	5.49 	&
  6.36 	&	9.72 	&	3.05 	&	14.85 	&	0.00 	&	2.20 	&	
  9.84 	&	7.83 	&	10.52 	&	3.83 	&	7.73 	&	0.00 	
  \\ \hline
  AC weight cuts	 & 1.83 	&	0.47 	&	18.23 	&	6.74 	&	0.00 	&	0.09 	&	
  0.00 	&	0.31 	&	13.77 	&	1.52 	&	0.00 	&	0.12 	&	
  20.17 	&	0.59 	&	19.88 	&	23.63 	&	0.08 	&	0.00 	
  \\ \hline
 CB sideband subtraction	 & 31.84 	&	0.0 	&	2.1 	&	8.81	&	0.88 	&	3.79 	& 
 4.34 	&	2.36	&	0.0 	& 1.67 	&	7.58	&	20.32 	&	
 77.31 	& 8.16	&	0.0 	& 2.49	&	6.33	&	13.28 
 \\ \hline
 Radiative correction		 & 3.87 	&	0.24 	&	0.00 	&	0.03 	&	0.00 	&	0.19 	&
 0.18 	&	0.28 	&	0.32 	&	0.54 	&	0.00 	&	0.12 	&	
 5.06 	&	0.49 	&	0.27 	&	0.01 	&	0.00 	&	0.00 	
 \\ \hline  \hline
Total   			 & 33.88 	&	11.38 	&	20.21 	& 11.28	&	8.96 	&	9.84	&
8.19 	&	13.18	&	14.96 	&	15.04	&	7.63	&	20.46	&
80.89 	&	11.84	&	23.35 	&	24.07	&	10.62 	&	13.95 	
\\ \hline
\end{tabular}}
\vspace{2.5cm}
\small
\caption{$A$-dependent transverse momentum broadening systematic effects and their contributions.}
\label{pTBroadA-SysTable}
\centering
\begin{tabular}{| c | c | c | c |}
\hline
\multirow{2}{*}{Systematic effect} & \multicolumn{3}{ c |}{Point-to-point Systematic Uncertainty ($\%$)} \\ \cline{2-4}
 & \hspace{0.25cm} Carbon \hspace{0.25cm} & \hspace{0.5cm} Iron \hspace{0.5cm} & Lead  \\ \hline
 Particle identification cuts	&	4.69	&	1.35	&	0.00
 \\ \hline
 Vertex Corrections	&	2.70	&	0.00	&	0.38	
 \\ \hline
 AC 6D map variables \& binning &	0.57	&	0.00	&	2.21
 \\ \hline
 AC weight cuts	&	3.40	&	0.00	&	7.07
 \\ \hline
 CB sideband subtraction	&	5.52	&	0.0	 &	2.87
 \\ \hline
 Radiative correction	&	0.04	&	0.17	&	0.00	
 \\ \hline \hline
Total				&	8.46	&	1.36	&	7.96	
\\ \hline
\end{tabular}
\end{table}
\end{turnpage}

\paragraph{\bf{SP.4~Tabulated Multiplicity Ratio and $p_T$-broadening Results}}
\label{Results}

This section contains the reported results in the last two figures of this manuscript, Figs.~\ref{Fig:mratioWithSystematics_shifted} and~\ref{Fig:pT2GIBUU}, detailed in Table~\ref{tab:mr} for all nuclei $z$-binned multiplicity ratios, and Table~\ref{tab:pt2z} (Table~\ref{tab:pt2A}) for all nuclei $z$-binned ($A$-dependent) transverse momentum broadening. In addition, the correlation between $z$ and the Feynman variable $x_F$ is illustrated in Fig.~\ref{fig:zvsxf} to support the discussion related to the separation between forward and backward fragmentation regions. Furthermore, the $z$-binned distributions, as well as AC-weighted averages, of the Bjorken scaling variable $x_B$ are shown in Fig.~\ref{fig:xB} and Table~\ref{tab:xBavr} to illustrate our kinematical coverage for any theoretical calculations aiming to describe our data.     


\begin{table}[h]
\caption{\label{tab:mr}Measured $\Lambda$ $z$-binned multiplicity ratios for all nuclei along with their total statistical and systematic (point-to-point and normalization uncertainties depicted in Fig.~\ref{Fig:mratioWithSystematics_shifted} added in quadrature) uncertainties.}
\centering
\hskip-1.0cm \begin{tabular}{c | c  c  c}
\hline \hline
\multirow{2}{*}{$z$ bin} & & $R^{A}_{\Lambda}$ $\pm$  Statistical $\pm$  Systematical Uncertainties &  \\ \cline{2-4}
 & Carbon & Iron & Lead\\ \hline
0.28 - 0.38 & 3.4256 $\pm$ 0.5319 $\pm$ 0.3004 & 5.7536 $\pm$ 0.5681 $\pm$ 1.2661 & 7.2363 $\pm$ 0.9997 $\pm$ 0.9893 \\
0.38 - 0.44 & 1.3447 $\pm$ 0.1603 $\pm$ 0.1628 & 1.9382 $\pm$ 0.1769 $\pm$ 0.3629 & 2.6378 $\pm$ 0.3405 $\pm$ 0.3863 \\
0.44 - 0.51 & 1.1084 $\pm$ 0.1205 $\pm$ 0.3299 & 2.0100 $\pm$ 0.1735 $\pm$ 0.3674 & 2.1293 $\pm$ 0.2316 $\pm$ 0.4987 \\
0.51 - 0.60 & 1.1498 $\pm$ 0.0883 $\pm$ 0.1400 & 1.2126 $\pm$ 0.0823 $\pm$ 0.1663 & 1.1857 $\pm$ 0.1057 $\pm$ 0.2659 \\
0.60 - 0.75 & 1.1174 $\pm$ 0.0756 $\pm$ 0.1519 & 0.9660 $\pm$ 0.0617 $\pm$ 0.1588 & 0.8910 $\pm$ 0.0759 $\pm$ 0.2364 \\
0.75 - 1.00 & 0.9506 $\pm$ 0.1011 $\pm$ 0.0741 & 0.6450 $\pm$ 0.0529 $\pm$ 0.1549 & 0.5622 $\pm$ 0.0621 $\pm$ 0.2096 \\ \hline \hline
\end{tabular}
\end{table}

\begin{table}[h]
\caption{\label{tab:pt2z}Measured $\Lambda$ $z$-binned $p_{T}$-broadening results for all nuclei with their total statistical and systematic (point-to-point and normalization uncertainties depicted in Fig.~\ref{Fig:pT2GIBUU} left added in quadrature) uncertainties.}
\centering
\hskip-1.0cm\begin{tabular}{c | c c c}
\hline \hline
\multirow{2}{*}{$z$ bin} & & $\Delta p^2_{T}$ (GeV$^2$) $\pm$  Statistical $\pm$  Systematical Uncertainties &  \\ \cline{2-4}
 &  Carbon &  Iron &  Lead \\ \hline
0.28 - 0.38 & 0.0003  $\pm$ 0.0143 $\pm$ 0.0015 & 0.0112  $\pm$ 0.0127 $\pm$ 0.0015 & -0.0072  $\pm$ 0.0151 $\pm$ 0.0060 \\
0.38 - 0.44 & 0.0259  $\pm$ 0.0160 $\pm$ 0.0033 & 0.0422  $\pm$ 0.0140 $\pm$ 0.0057 & 0.0592  $\pm$ 0.0171 $\pm$ 0.0071 \\
0.44 - 0.51 & 0.0648  $\pm$ 0.0174 $\pm$ 0.0132 & 0.0894  $\pm$ 0.0147 $\pm$ 0.0134 & 0.0613  $\pm$ 0.0174 $\pm$ 0.0144 \\
0.51 - 0.60 & 0.1317  $\pm$ 0.0165 $\pm$ 0.0149 & 0.2120  $\pm$ 0.0168 $\pm$ 0.0319 & 0.2007  $\pm$ 0.0211 $\pm$ 0.0483 \\
0.60 - 0.75 & 0.1879  $\pm$ 0.0225 $\pm$ 0.0169 & 0.2591  $\pm$ 0.0218 $\pm$ 0.0198 & 0.3140  $\pm$ 0.0295 $\pm$ 0.0334 \\
0.75 - 1.00 & 0.1145  $\pm$ 0.0157 $\pm$ 0.0114 & 0.1381  $\pm$ 0.0149 $\pm$ 0.0283 & 0.1788  $\pm$ 0.0209 $\pm$ 0.0250 \\  \hline \hline
\end{tabular}
\end{table}
    
\begin{table}[h]
\caption{\label{tab:pt2A}Measured $\Lambda$ $A$-dependent $p_{T}$-broadening results for all nuclei along with their total statistical and systematic (point-to-point and normalization uncertainties depicted in Fig.~\ref{Fig:pT2GIBUU} right added in quadrature) uncertainties.}
\centering
\begin{tabular}{c|c}
\hline \hline
$A$ & $\Delta p^2_{T}$ (GeV$^2$) $\pm$  Statistical $\pm$  Systematical Uncertainties \\ \hline
 Carbon & 0.0952 $\pm$ 0.0272 $\pm$ 0.0082 \\
 Iron & 0.1404 $\pm$ 0.0376 $\pm$ 0.0024 \\
 Lead & 0.1823 $\pm$ 0.0451 $\pm$ 0.0146 \\   \hline \hline
\end{tabular}
\end{table}

\begin{table}[h]
\small
\caption{$z$-binned $x_B$ AC-weighted averages for all nuclei.}
\label{tab:xBavr}
\centering
\begin{tabular}{| c | c | c | c | c |}
\hline
\multirow{2}{*}{$z$ bin} & \multicolumn{4}{ c |}{$x_B$ AC-weighted Average} \\ \cline{2-5}
 & \hspace{0.25cm} LD2 \hspace{0.25cm} & \hspace{0.5cm}  Carbon \hspace{0.25cm} & \hspace{0.5cm} Iron \hspace{0.5cm} & Lead  \\ \hline
 0.28 - 0.38 &	0.2176	& 0.2395 & 0.2401 & 0.2391
 \\ \hline
 0.38 - 0.44	& 0.2529 &	0.2574 & 0.2623 & 0.2551
 \\ \hline
 0.44 - 0.51 & 0.2585 & 0.2665 & 0.2701 & 0.2724
 \\ \hline
 0.51 - 0.60	& 0.2661 &0.2699 & 0.2697 & 0.2703
 \\ \hline
 0.60 - 0.75	& 0.2756 & 0.2648 & 0.2667 & 0.2646
 \\ \hline
 0.75 - 1.00	& 0.2921 &	0.2858 & 0.2821 & 0.2817
 \\ \hline 
\end{tabular}
\end{table}

\end{document}